# Exit, Voice and Political Change:
# Evidence from Swedish Mass Migration to the United States
# A Comment[*]

Per Pettersson-Lidbom[#]

This version: January 12, 2022

**Abstract**

In this comment, I revisit the question raised in Karadja and Prawitz (2019) concerning a causal relationship between mass emigration and long-run political outcomes. I discuss a number of potential problems with their instrumental variable analysis. First, there are at least three reasons why their instrument violates the exclusion restriction: (i) failing to control for internal migration, (ii) insufficient control for confounders correlated with their instrument, and (iii) emigration measured with a nonclassical measurement error. Second, I also discuss two problems with the statistical inference, both of which indicate that the instrument does not fulfill the relevance condition, i.e., the instrument is not sufficiently correlated with the endogenous variable emigration. Correcting for any of these problems reveals that there is no relationship between emigration and political outcomes.

[*] I am grateful to Erik Prawitz for providing a file with the names of the municipalities in order to match my variable to their JPE data set. I am also grateful to Erik Prawitz. Mounir Karadja, Björn Tyrefors and David Strömberg for useful discussions.
[#] Department of Economics, Stockholm University, and Research Institute of Industrial Economics (IFN), E-mail: pp@ne.su.se

# 1. Introduction

Karadja and Prawitz (2019) (henceforth KP) estimate the effect of emigration (i.e., external migration) on long-run political outcomes using historical data from Sweden for the period 1867-1920. KP estimates a cross-sectional regression where the unit of observations is a municipality (i.e., 2359 geographical units). KP uses an instrumental variable (IV) approach where the instrumental variable for emigration is an interaction between a weather phenomenon, *Shocks*, i.e., the number of frost shocks measured at the weather station level (i.e., 32 stations), and the geographical distance between the closest port of emigration and the municipality of residence, *Port*. They argue that their instrument is exogenous conditional on a set of control variables, which is formally expressed in their paper on page 1886 as

(1)  $\mathrm{E}[\varepsilon_{mct} | Shocks \times Port_{mc}, Shocks_{mc}, Port_{mc}, \Phi_c, X'_{mc}] = 0.$

Thus, KP's identifying assumption (i.e., the exclusion restriction) is that their instrument, i.e., *Shocks×Ports*, only affects emigration *conditional* on a set of control variables, i.e., *Shocks*, *Port*, $\Phi_c$ (24 county fixed effects), and other pretreatment variables as denoted by $X'$.

In this paper, I discuss a number of potential problems with their IV analysis.[1] First, there are at least three reasons why their instrument violates the exclusion restriction: (i) the assumption that internal migration is not affected by the instrument is incorrect, (ii) insufficient control for confounders correlated with their instrument, and (iii) emigration is measured with a nonclassical measurement error. Second, there is also a problem with the statistical inference in their IV approach. One problem concerns that their reported standard errors and *F*-statistics are not valid in settings with clustered data and highly leveraged observations (e.g., Young (2021)). Another problem is that the reported first-stage statistics are wrong, i.e., to large, since it does not consider that the KP uses a constructed instrument (e.g., Hull (2017)) The conclusion from this comment is that KP's analysis is not credible due to violation of the exclusion restriction and problems related to weak instruments.

The rest of the paper is structured as follows. In Section 2, I discuss three reasons why the instrument in KP violates the exclusion restriction. In Section 3, I discuss two problems with the inference. Section 4 concludes.

---

[1] My critique also concerns Andersson, Karadja and Prawitz (forthcoming) which use the same identification strategy. My critique concerning the violation of the exclusion restriction has previously been discussed in a working paper, i.e., Pettersson-Lidbom (2020), which has been commented by Kardaja and Prawitz (2020). However, Andersson, Karadja and Prawitz (forthcoming) do not discuss this critique.

## 2. Violations of the exclusion restriction

In this section, I discuss three reasons why KP's instrument violates the exclusion restriction expressed in equation (1), i.e., (i) failing to control for internal migration, (ii) insufficient controls for confounders correlated with the instrument, and (iii) emigration is measured with a nonclassical measurement error.

The first two issues can be thought of as different types of omitted variable bias (OVB) problems. Thus, failing to control for these variables makes the population error term in equation (1) above $\varepsilon$ correlated with the instrument *Shocks×Ports*. The third issue is conceptually different since it has to do with bias in the instrumental variable approach as caused by nonclassical measurement errors in the explanatory variable. Nonetheless, it still leads to a violation of the exclusion restriction.

### 2.1 Failing to control for internal migration

KP (implicitly) assumes that their instrument does not affect internal migration but only emigration (i.e., external migration) since internal migration is omitted from their regression specifications, i.e., equations 1-3.[2] However, if the instrument also affects internal migration,[3] then the exclusion restrictions will be violated. Indeed, KP shows that internal migration is affected by the instrument (see Column 1 in Table 8 in KP). Thus, this result shows that their instrument violates the exclusion restriction. However, this issue is not discussed in the paper. To solve this problem, internal migration must be controlled for in KP's specifications. Moreover, an additional instrument for internal migration is also necessary for identification since internal migration is an endogenous variable such as emigration.[4]

### 2.2 Insufficient control for confounders correlated with the instrument

In this section, I will discuss another omitted variable problem in KP's analysis, namely, that they fail to sufficiently control for confounders correlated with the instrument. The issue has to do with that KP's instrument is based on geographic variation, i.e., the interaction between *Shocks* and *Port,* and where *Shocks* and *Ports* are both geographical variables. In such a case, it is crucial to flexible control for geographical characteristics or any other factor associated with these characteristics as to avoid OVB. Otherwise the instrument will not be conditionally exogenous as explained in the following.

---

[2] For example, equation (1) in KP states that $y_{cmt} = \beta Emigration_{cmt} + \Phi_S + X'_{mc}\beta_X + \eta_{cmt}$. Thus, a more general specification would also include internal migration as an explanatory variable, i.e., $y_{cmt} = \beta Emigration_{cmt} + \pi Internal\_migration + \Phi_S + X'_{mc}\beta_X + \eta_{cmt}$.
[3] Is seems implausible to think that that the instrument does not affect internal migration. For example, if an individual living in the north part of Sweden (i.e., with a long distance to ports) experience a frost shock, then she/he might decide move, but not necessarily abroad.
[4] Another solution is to define the parameter of interest as the effect of all types of migration, i.e., internal plus external, on long-run political outcomes. Thus, $y_{cmt} = \beta Migration + \Phi_S + X'_{mc}\beta_X + \eta_{cmt}$, where Migration= Emigration + Internal migration. In this case, it is sufficient to have only one instrument.

KP writes "An important feature of our identification strategy is that we control for the direct effects of frost shocks and port proximity in (4). This is beneficial because studies that use weather shocks as instruments are typically marred by the problem that weather may simultaneously affect many variables (Giuliano and Spilimbergo 2014; Sarsons 2015). In our setting, locations that experience more severe frost shocks may obtain weaker government finances, worse public health, or other features that can affect our outcomes without going through emigration."

Thus, the central idea of KP's identification approach is that they can control for the frost shock, *Shocks*, and port proximity, *Port*, while the interaction between these two variables, *Shocks×Port*, is assumed to be a valid instrumental variable. Thus, KP assumes that their instrument is exogenous *conditional* upon the control variables *Shocks* and *Port*. However, there are two problems with this approach of controlling for the confounding effects of the frost shocks using only a single control variable, *Shocks*.

The first problem is that KP has imposed that the control variable *Shocks* only has a *linear* effect on the outcome of interest even though that this assumption can be easily relaxed. It is however unlikely that the frost shocks only affect the outcome linearly since KP the variable *Shocks* is so highly geographically clustered (see Figure 4 in KP), Thus, it is therefore crucial to control completely for unobserved geographic heterogeneity at the shock level as to avoid OVB.

The second problem is also similar since KP does not control for unobserved geographical characteristics at the appropriate level, i.e., weather station level. As a result, their instrument may be correlated with unobserved characteristics across the spatial areas where frost shocks occur, i.e., at the weather station level, even after conditioning on the control variable *Shocks*.

Starting with the problem of imposing linearity on the variable *Shocks*, it is possible to completely relax this assumption by estimating a model with a full set of indicators for each level of the frost shocks.[5] Since the control variable *Shocks* only takes 12 distinct values, it is sufficient to include 11 dummy variables to completely saturate the model. An $F$-test strongly rejects the assumption of linearity (i.e., $F(11, 31) = 7.1$ and $\text{Prob} > F = 0.0000$) in the first-stage relationship.[6] Moreover, the estimated first-stage effect is 0.0226 with a standard error of 0.0209 when relaxing the linearity assumption. Thus, this estimate is much smaller than KP's reported first-stage estimate of 0.0621 and is not statistically significantly different from zero. Thus, allowing the effect of frost shocks on the outcome of interest to be nonlinear shows that there is no first-stage effect.[7]

---

[5] In a Web Appendix (Table B11), KP estimate models with different polynomials of *Shocks* and *Port*. However, these polynomial regressions still impose strong functional form assumptions. More important, the variable *Shocks* cannot be accurately approximated by a *continuous* function, such as polynomials or splines, since it is a *discrete* variable, measuring the number of frost shocks, that only takes 12 values.

Turning to the problem that KP's instrument may be correlated with unobserved heterogeneity at the weather station level, it is possible to address this issue by including a full set of weather station fixed effects since KP's instrument is an interaction between frost shocks and port proximity.[8] In other words, even when spatial fixed effects at the weather station level are controlled for, there will still be variation in KP's instrument within weather stations. Nonetheless, KP omits fixed effects at the weather station level in their analysis. Instead, they include county-specific effects, i.e., $\Phi_c$ in equation (1) above. However, these geographical fixed effects cannot control for confounders at the level of the weather shocks, since counties comprise an administrative level of the central government consisting of 24 geographical areas, and each individual county fixed effect in KP's analysis will map into municipal data from approximately four weather stations, on average.[9] As a result, county fixed effects are therefore not adequate controls for unobserved heterogeneity at the weather station level. Moreover, the choice of controlling for fixed effects at the county level seems more or less arbitrary since these effects are only partly related to the other key control variable *Port*, which measures the geographical distance to ports. As a result, county fixed effects are unlikely to be sufficient controls for the effect of distance to ports not captured by the single control variable *Port*, such as nonlinear geographical effects and other time-invariant geographic characteristics correlated with *Port*.[10]

---

[6] This specification corresponds to the one in Column 4 in Table 3 in KP, i.e., with a first-stage estimate of 0.621.
[7] However, the reduced form effect is statistically significant, i.e., 0.00184 with a standard error of 0.00086. Thus, this finding also suggest that the exclusion restriction is violated as further discussed below.
[8] Dell et al. (2014) discuss the crucial importance of controlling for spatial (area) fixed effects when using a "weather-shock" approach in a panel data setting, i.e., using an exogenous source of variation in climatic variables over time within a given spatial unit. Indeed, they write that "the weather-shock approach has strong identification properties. The fixed effects for the spatial areas, $\mu_i$, absorb fixed spatial characteristics, whether observed or unobserved, disentangling the shock from many possible sources of omitted variable bias."
Although Dell et al. (2014) discuss panel data applications, it is also possible to include fixed effects for the spatial areas in which the weather shocks occur even in a cross-sectional approach, as in KP's analysis, as long as the weather shock is interacted with some other variable that varies within the spatial areas.
[9] There are a total of 32 weather stations in KP's data.

Controlling for weather station fixed effects, the first-stage effect is 0.0189 with a standard error of 0.0464 in the specification with the full set of controls.[11] Thus, controlling for weather station fixed effects shows that there is no first-stage relationship when unobserved heterogeneity at the weather station level is considered.[12]

**2.3 Emigration is measured with a nonclassical measurement error**

In this section, I will discuss the problem that emigration is measured with a nonclassical measurement error; that is, the variable of interest (emigration) and its measurement error are *not* uncorrelated, and the expected value of the mismeasured variable is *not* equal to the expected value of the true measure. As a result, KP's instrumental variable approach will be biased. I will also present a solution to the measurement error problem.

It is well known that the Swedish emigration statistics during the $19^{th}$ century and early $20^{th}$ century are unreliable due to the severe underreporting of emigrants. This has been documented and discussed, for example, by Statistics Sweden in an official report from 1887, Emigrationsutredningen (1909, p. 593), Johansson (1976), Odén (1964, 1971), Ahlberg (1976), Eriksson (1969), Hofsten and Lundström (1976), and Vernersson Wiberg (2016). These studies show not only that the emigration to the U.S. was severely underreported but also that the emigration to other countries within Europe (e.g., Denmark and Germany) was even more underreported.

It is noteworthy that the studies discussing the problems with the Swedish emigration statistics are not cited in KP.[13] As a result, the discussion in KP that the emigration data are reliable on p. 1876 is questionable.[14] Specifically, the claim "it is possible to ascertain their

---

[10] KP write (p. 1885) that "By including county fixed effects and using proximity in logarithms (rather than levels), the identifying variation does not disproportionately rely on northern counties." However, a much better approach is to order all 2,359 municipalities based on the distance to ports and define geographical groups accordingly. In this way, it is possible to control much more convincingly for the factors related to distance to ports that seem to concern KP.
[11] This specification corresponds to the one in Column 4 in Table 3 in KP, i.e., with a first-stage estimate of 0.621.
[12] The estimated clustered standard errors also become significantly larger than those reported by KP (e.g., 0.046 vs. 0.015), suggesting that KP's clustered standard errors are likely to be biased downward because there is a correlation of the errors within weather stations that is not properly dealt with by only using clustered standard errors when there are only a limited number of clusters (32) and very unequal cluster sizes (i.e., the range is from 2 to 311).
[13] This literature should be familiar to KP since I provided references to this work already in 2015 when I suggested that they must address the problem with measurement errors in the Swedish emigration statistics.
[14] KP's claim that their data sets encompass "the universe of registered emigrants during the Age of Mass Migration" is also incorrect since their parish data is estimated to contain only 75% of all emigrants. Data from

[parish reports and ship passenger lists] accuracy by cross-checking the two sources" is erroneous since parish records reported emigration to all countries, while ship passenger lists essentially recorded only emigration to the United States. Indeed, Eriksson (1969) finds that the overlap of individuals between these two sources is only 44%. Part of this discrepancy is due to parish records registering only individuals with a change-of-address certificate.[15] Thus, KP cannot solve the underreporting problem by using a "single emigration variable defined the maximum of either the church book or passenger list data each year" since there will be a very large number of missing emigrants.[16] Moreover, even unifying the two data sets would be insufficient since there would still be a large number of emigrants who are not recorded in either of these sources, i.e., those who did not apply for a change-of-address certificate and who emigrated to countries other than the U.S. A similar point is also made in Johansson (1976) and Odén (1971).

Most importantly, I estimated that KP's emigration variable only includes at most 73% of all emigrants during the period 1860-1920.[17] As a result of this large underreporting of emigration, the KP instrumental variable approach will be inconsistent due to this type of nonclassical measurement error.

To formally illustrate the measurement error problem in KP and how it can be solved, let $X^*$ denote the true emigration. The population regression of interest in KP's analysis can now be expressed as

(2) $\qquad Y_i = \alpha + \beta X_i^* + u_i,$

where $Y_i$ is some political outcome in municipality $i$, and $X_i^*$ is the *true* total sum of emigrants who emigrated (i.e., moved *outside* Sweden) from municipality $i$ during the period 1867-1920. KP uses an instrumental variable approach in which they replace the true value of $X_i^*$ with an error ridden measure, $X_i$, as noted above. Then, they assume that their instrumental variable, $Z_i$, is uncorrelated with both the population error term $u_i$ and the reporting error $e_i = X_i - X_i^*$.

---

a number of parishes is also missing in their data (see link https://emiweb.se/?services=emigranter-i-svenska-kyrkbocker/%20).

[15] This problem has been regarded as the chief explanation of the discrepancy between actual and recorded emigration, Johansson (1976).

[16] Importantly, KP lack data from the church books after 1895.

[17] This calculation is partly based on the official statistics (https://www.scb.se/en/finding-statistics/statistics-by-subject-area/population/population-composition/population-statistics/pong/tables-and-graphs/yearly-statistics--the-whole-country/population-and-population-changes/), i.e., those with a change-of-address certificate, which recorded 1.3 million emigrants during the period 1860-1920. I have also estimated that a minimum of 0.2 million emigrants were not recorded during this period due to various sources of errors discussed by Johansson (1976) and Eriksson (1969), among others. Thus, at least 1.5 million emigrated from Sweden during the period 1860-1920. Consequently, a minimum of 0.4 million emigrants are missing from KP's data since it only includes 1.1 million emigrants.

However, because KP replaces the true value in the equation with the error-ridden value, the instrument variable estimator is *not* consistent since the probability limit of the instrumental variable estimator can now be written as

$$(3)\quad plim\ \beta^{IV} = \frac{Cov(Y,Z)}{Cov(X,Z)} = \frac{Cov(\beta X^* + u, Z)}{Cov(X^* + e, Z)} = \frac{\beta Cov(X^*, Z) + Cov(u, Z)}{Cov(X^*, Z) + Cov(e, Z)}$$

and, due to nonclassical measurement errors, i.e., $Cov(e, X^*) \neq 0$, the instrument will also be correlated with the reporting error, i.e., $Cov(e, Z) \neq 0$. Thus, $plim\ \beta^{IV} \neq \beta$ even if $Cov(u, Z) = 0$ holds. In fact, equation (3) shows that the estimate from the instrumental variable method will be biased upward if the measurement error is negatively correlated with the true value, i.e., $Cov(e, Z) < 0$, and biased downward if $Cov(e, Z) > 0$.[18, 19] It is, however, *a priori* difficult to assess the direction of this bias since the true measure of emigration is not known and that the instrument is an interaction variable, *Shocks×Port*. Nonetheless, there is still a bias in the KP instrumental variable approach due to a nonclassical measurement error.

The inconsistency problem in PK's instrumental variable approach can, however, be solved by finding a measure of emigration that has classical measurement errors instead of nonclassical errors.[20] In fact, the registered total outmigration, i.e., the sum of the true emigration, $X_i^*$, and the true internal migration, $I_i^*$, fulfills the classical assumption since internal migration is *excluded* from the explanatory variables in KP's population regression model. In other words, KP have (implicitly) assumed that their instrument $Z_i$ is unrelated to internal migration $I^*$ since it is subsumed in the population error term.[21] As a result, it is possible to replace $X_i^*$ in equation (2) with total outmigration, i.e., $X_i^* + I_i^*$, and still obtain a consistent estimate of $\beta$ since $Cov(I^*, Z)$ is assumed to be zero in KP's analysis.

I have collected data on total outmigration from the Swedish National Archives for the period 1860-1950 as part of my ERC-financed historical database project. With these data, it is possible to assess to what extent the results in KP are affected by the problem of underreporting emigrants.[22] Interestingly, the reported emigration used by KP only makes up, on average, 8% of the total outmigration (the median value is 6%) during the period 1867-1920, and the share is never larger than 20% for any individual year. Thus, this value must be considered a very low share given the very large Swedish emigration during this period since it has been estimated that at least 1.5 million people emigrated out of an average population of only 4.8 million during the period 1860-1920. Thus, this finding further underscores the problem of underreporting of the Swedish emigration in KP's data.

---

[18] The expression of the IV estimator in equation (2) does not consider that there may be errors in both *Y* and *Z*, that other included variables may be measured with errors, and that all these errors may be corelated with each other. In such a general model, it is virtually impossible to sign the direction of the bias of the IV estimator. For example, KP's instrument can be measured with an error that is correlated with *e* which then must be taken into consideration.

[19] See also Bound et al. (2001) for a general treatment of measurement errors.

[20] Bound et al (2001, p. 3729) write that "strategies for obtaining consistent estimates of the parameters of interest work if the measurement error is classical, but do not, in general do so otherwise.

[21] Indeed, KP treat internal migration *I\** as an additional outcome variable *Y* in their instrumental variable approach in Column 1 in Table 8. Thus, KP therefore have assumed the following causal chain: $Z \rightarrow X^* \rightarrow Y$, i.e., the instrument *Z* only has an indirect effect, which only goes through *X\**, on the outcome *Y*, i.e., $Cov(I^*, Z) = Cov(u, Z) = 0$. If this exclusion restriction is wrong, i.e., that *Z* has a *direct* effect on both *X\** and *I\**, then two valid instruments are required for identification, i.e., one for *X\** and another for *I\**. This may be another reason why KP's empirical analysis is flawed.

Turning to the result of the solution of the measurement error problem, the first-stage estimate with total outmigration is 0.0100 with a standard error of 0.0076, while KP's estimate is 0.0621.[23] Thus, KP's estimate is biased upward with a factor of more than 6. This first-stage estimate with total outmigration is also precisely estimated to be zero since it can rule out a first-stage effect larger than 0.0255. Consequently, there is no first-stage relationship in KP's analysis when correcting for the problem of underreporting.

## 2.4. Further tests of violation of the exclusion restrictions

In this section, I present some additional test of the potential violation of the exclusion restriction.

One way to check the plausibility of an instrumental variable approach is to check whether the first-stage estimate (FS) and reduced form (RF) estimate are sensitive to changes in the sample/population/functional form assumptions (e.g., Angrist and Pischke (2009) and Bazzi and Clemens (2013)).

One such check is to probe the sensitivity to the very strong assumption of a linear relationship between the IV and the control variable distance to ports, *Port*, which is one of the variables in KP's constructed instrument, i.e., *Shocks×Port*. Importantly, *Port* is not as good as randomly assigned as *Shocks* but instead a fixed (time-invariant) characteristic. As discussed above, KP assumes that the instrument is valid by conditioning only on the linear term, *Port*. This assumption is problematic since frost shocks and distance to ports are almost perfectly correlated for local governments situated in the northern part of Sweden (i.e., north of Stockholm). As a result, these observations do not provide much independent information in the IV approach except for the imposed functional form. Thus, it is important to understand how important the functional form assumption is for identification in KP.

---

[22] In this footnote, I describe how my variable, the cumulative sum of total outmigration for the period 1867-1920, was merged to KP's data set that I downloaded from JPE's homepage. I discovered that KP's data files do not include the names of the geographical areas (i.e., municipalities) but only a variable running from 1 to 2,359. Thus, I had to ask KP to send me this information. After some work, I was able to match 2,330 out of the 2,359 municipalities by using the code developed by Riksarkivet.

[23] This specification corresponds to the one in Column 4 in Table 3 in KP, i.e., with a first-stage estimate of 0.621.

To check the sensitivity to this assumption, I estimate exactly the same FS and RF models as KP but for different distances to the port. If the instrument fulfills the exclusion restriction, we would expect that the FS and RF estimates should be relatively unchanged irrespective of the distance to port. On the other hand, if the FS and RF estimates switch signs or have very different magnitudes depending on the distance to port, then this strongly suggests that the IV approach is flawed.

In Table 1, I have estimated exactly the same first-stage and reduced form regressions as preferred by KP but where I include different municipalities depending on the (log) distance to port, *lproxemiport*. In column 1, I have included the estimates from KP (2019) for ease of comparison with my results. In column 2, I only include those municipalities that are closest to port, i.e., those with *lproxemiport* >2 as measured according to their variable. There are 83 such municipalities. The estimated first-stage is then 0.332, and the corresponding reduced-form effect is -0.068. Thus, the estimates are very different (both in magnitude and sign) from the KP result shown in Column 1. In Column 3, I add those municipalities with *lproxemiport* >1.5; in column 4, I add those with *lproxemiport* >1; in column 5, I add those with *lproxemiport* >0.5, …; and finally, in Column 9, I add those with *lproxemiport* <-1.5. These specifications thus include an increasing number of municipalities with an increasing distance to emigration ports, i.e., 83, 177, 338, 683, 1099, 1527, 2070, 2291.

The results in Table 1 show that the FS estimate not only switches sign but is of very different magnitudes, ranging from -0.360 to 0.332. I find similar results for the RF since the estimates range from -0.068 to 0.021. Moreover, the relationship between the estimate of the FS and RF is erratic and does not make any sense.[24] It is only in the last two specifications when municipalities furthest way from emigration ports are also included in the estimation (i.e., when at least 2,070 of all the 2,359 municipalities) that the estimates are broadly similar to KP (2019). Again, these results reveal that the exclusion restriction is violated.

---

[24] This is also known as visual IV as discussed by Angrist and Pischke (2009). For example, Angrist and Kreuger (1991) compare the relationship between FS and RF.

Table 1. First-stage and reduced-form estimates depending on the distance to port

| | (1) | (2) | (3) | (4) | (5) | (6) | (7) | (8) | (9) |
|---|---|---|---|---|---|---|---|---|---|
| Distance to port: *lproxemiport* | | >2 | >1.5 | >1 | >0.5 | >0 | >-0.5 | >-1 | >-1.5 |
| First-stage effect | .062*** | .332 | -.072 | .263 | -.360* | -.0052 | -.0046 | .091*** | .063*** |
| | (.015) | (.75) | (.116) | (.131) | (.157) | (.106) | (.066) | (.024) | (.017) |
| Reduced form effect | .0017*** | -.068*** | .004 | .0125*** | .0129* | -.0023 | -.0014 | .0021*** | .0015*** |
| | (.0004) | (.011) | (.004) | (.002) | (.006) | (.005) | (.004) | (.0005) | (.0004) |
| | | | | | | | | | |
| Number of municipalities in the sample | 2,359 | 83 | 177 | 338 | 683 | 1,099 | 1,527 | 2,070 | 2,291 |

Note: Coefficients significantly different from zero are denoted by the following system: *10 percent, **5 percent, and ***1 percent

Another way of detecting whether the exclusion restriction is violated is to check the sensitivity of the FS and RF to various alterations of the included control variables.[25] In the best-case scenario, both the RF and FS should be insensitive to the control variables included. Otherwise, one must take a stance on whether a variable is either a good or a bad control variable (e.g., Cinelli et al. (2021)). For example, Cinelli et al. (2021) show that a pretreatment characteristic should not be controlled if it is a collider variable, i.e., a variable that is causally influenced by two or more variables.

I will use three different specifications to illustrate that the results in KP are highly sensitive to the included control variables. Table 2 presents the results from first-stage estimates (Panel A) and reduced-form effects on labor organizations (Panel B).

Table 2. First-stage and reduced-form results from three different specifications

|  | (1) | (2) | (3) |
|---|---|---|---|
| Panel A. First-stage effect | | | |
| KP's instrument | 0.189*** | 0.063*** | 0.042 |
|  | (0.041) | (0.016) | (0.032) |
| Panel B. Reduced form effect | | | |
| KP's instrument | -0.0002 | 0.0014*** | -0.0015** |
|  | (0.0006) | (0.0004) | (0.0006) |

Note: Coefficients significantly different from zero are denoted by the following system: *10 percent, **5 percent, and ***1 percent.

Column 1 reports the results from a "bare bones" specification that includes the key control variables used by KP (i.e., Shocks, Port, the initial population size in 1865) with the exception of county fixed effects. Column 1 shows that the first-stage effect is very large, i.e., 0.189, compared to KP's reported first-stage effect of approximately 0.06. However, the reduced form effect is nonetheless very small and even negative, i.e., -0.0002, and thus very different from KP's results, where they report a reduced form effect in the range 0.0014-0.0017.

Column 2 reports the results from the addition of county fixed effects to the bare-bones specification in Column 1. Now, the first-stage effect is reduced to 0.063, while the reduced-form effect becomes positive and much larger, i.e., 0.0014, and statistically significant. Thus, it is noteworthy that a positive and statistically significant reduced-form effect only appears when controlling for county fixed effects.

---

[25] Karadja and Prawitz (2020) also conduct a similar exercise in their Table 2. They provide evidence that the exclusion restriction does not hold since the reduced form is insensitive to the corresponding first-stage estimate. Indeed, they find a reduced form effect of approximately 0.0016 independent of the size of the first-stage estimate which ranges from 0.0243 to 0.0632. Nonetheless, they still argue that their instrument is valid.

Column 3 shows the results when weather station fixed effects, instead of county fixed effects, are added to the bare-bone specifications in Column 1. Now, the first-stage effect decreases further, to 0.042, and is no longer statistically significant. However, the reduced form effect becomes negative, -0.0015, and is statistically significant. Thus, controlling for weather station fixed effects led to exactly the opposite reduced-form result compared to the specification with county fixed effects in Column 2.

Taken together, the results presented in Tables 1 and 2 strongly suggest that there does not exist a causal relationship between emigration and labor movements since there is no consistent relationship between the first-stage and reduced-form results in KP's analysis.

## 3. Inference problems

In this section, I describe two problems with the statistical inference in PK's analysis, both of which reveal that the instrument does not fulfill the relevance condition, i.e., that the instrument is not sufficiently correlated with the endogenous variable emigration.

### 3.1 High leverage observations and clustered data

The first problem is that the data are highly clustered, i.e., the variable *Shocks* only takes 12 distinct values since the number of frost shocks ranges from 0 to 11. Figure 4 in their paper shows a map of how these 11 frost shocks are geographically clustered across Sweden.[26] Young (2021) shows that in this type of setting with highly clustered data., the usual IV standard errors produced by Stata are susceptible to high leverage observations,[27] particularly with clustered and robust standard errors. Indeed, he argues that "statistically significant IV results generally depend upon only one or two observations or clusters". He proposes dropping one cluster at a time ("delete-one-sensitivity") to check whether the statistical inference is reliable. Specifically, he argues that "delete-one sensitivity, of *t*-statistics not coefficients, highlights the degree to which significant results depend upon sensitive coefficient and standard error estimates".

Applying the delete-one sensitivity approach to KP's IV analysis shows that dropping the cluster with three frost shocks produces a second-stage z statistic of 1.37 with an associated

---

[26] KP cluster their standard errors at the weather station level consisting of 32 clusters. However, since the instrument is constructed partly based on the number of the frost shocks, the standard errors should arguably be clustered at the *Shocks* level, which consists of 12 clusters. Thus, this another reason why the KP's statistical inference is wrong.

[27] The standard errors produced by Stata is based on that the normal approximation to the distribution of the reduced-form and first-stage coefficients is accurate. However, Young (2021) finds that normal approximation is unreasonable in settings with high leverage observations and clustered data.

P value of 0.17. Moreover, the robust first-stage $F$-statistic is equal to 2.89 when dropping the cluster with 3 frost shocks. Thus, the first-stage $F$ is now considerably below the rule of thumb of 10 of not having a problem with weak instruments.[28] Thus, the delete-one-delete-one sensitivity approach suggests that KP's statistical analysis is unreliable due to high leverage observations and clustered data.

### 3.2 Constructed instrument

The second problem is that KP's instrument, *Shocks×Port*, has been constructed and not estimated. As a result, the reported first-stage $F$-statistic does not take that issue into account. As discussed previously, KP constructs their single instrument by first creating a single variable, *Shocks*, i.e., the number of frost shocks, which is then interacted with the distance to port, *Port*. However, this model will not produce the correct $F$-statistics since the dimensionality of the underlying variation in the instrument is not one, as discussed by Hull (2017). Instead, the underlying variation should be based on a model with mutually exclusive and exhaustive binary variables of the constructed instrument. Thus, *Port* should be interacted with 11 binary variables where each binary variable corresponds to a particular value of *Shocks*.[29] Thus, this specification gives rise to 11 instruments instead of one. Hull (2017) shows that, as a rule of thumb, the true $F$ statistics are approximately $N$ times smaller than the reported one in the homoscedastic case, where $N$ is equal to the number of binary instruments. For example, KP reports in Column 6 in Table 5 an $F$-statistics of 17.19. However, the (homoscedastic) first-stage $F$ is only 4.2 (i.e., 46.655/11) when using Hull's degrees-of-freedom correction.[30] Thus, KP's statistical analysis is not valid due to a problem with weak instruments.[31]

## 3. Conclusion

In this paper, I revisit the question raised in Karadja and Prawitz (2019) concerning a causal relationship between mass emigration and long-run political outcomes. I find that their instrumental variable approach is invalid due to problems with both exogeneity and relevance of the instrument. Correcting for any of these problems reveals that there is no relationship between emigration and political outcomes.

---

[28] There is a recent literature that argues that the rule of thumb must be considerably larger than 10 to avoid the problem of weak instruments (e.g., Lee et al. (2021) and Keane and Neal (2021)).

[29] Moreover, by leveraging all 11 instruments makes it possible to inspect the individual first-stage effects to see whether the point estimates make sense. We would expect that the size of the first-stage estimate would increase monotonically with the number of frost shocks if KP's instrument is valid. However, this is not the case which also suggests that the instrument is not valid.

[30] I have also estimated that the effective $F$ statistic is 4.637 using Montiel Olea and Pflueger (2013) approach. This test for weak instruments also indicates that KP has a problem with weak instruments since the critical value is 30.574 when the maximal relative IV bias is 5%.

[31] See Andrews at al. (2019) for a state-of-the-art survey on weak instruments.